# Polariton-mediated binding of anti-aligned dipolar excitons

Haifeng Kang[1], Quanbing Guo[1,2*], Tianyi Zhou[2], Shun Feng[3], Wei Dai[1,2], Kenji Watanabe[4], Takashi Taniguchi[5], Hongxing Xu[1,2], Ting Yu[1,2*], Xiaoze Liu[1,2,6*]

[1] Key Laboratory of Artificial Micro/Nano Structure of Ministry of Education, School of Physics and Technology, Wuhan University, Wuhan, P. R. China
[2] Wuhan Institute of Quantum Technology, Wuhan, P. R. China
[3] Institute of Electrical and Microengineering, École Polytechnique Fédérale de Lausanne (EPFL), Lausanne, Switzerland
[4] Research Center for Functional Materials, National Institute for Materials Science, Ibaraki, Japan
[5] International Center for Materials Anorthite, National Institute for Materials Science, Ibaraki, Japan
[6] Wuhan National High Magnetic Field Center, Huazhong University of Science & Technology, Wuhan, 430074, P. R. China

## Abstract

Interacting bosonic quasiparticles are the cornerstone for exploring many-body physics and nonlinear quantum phenomena in correlated light-matter systems. Strongly interacting dipolar excitons in van der Waals heterostructures have attracted significant interest due to their out-of-plane electric dipole moments and high tunability via the quantum-confined Stark effect (QCSE). However, leveraging these tunable dipolar excitons in strongly coupled exciton-photon systems to explore exotic many-body physics and macroscopic quantum phenomena remains experimentally elusive. Here, we report the strong coupling of dipolar excitons in a gated bilayer $MoS_2$ device integrated with a one-dimensional photonic crystal hosting bound-states-in-continuum (BIC). The resulting polaritons hybridize cavity photons with a coherent superposition of two electrically tunable anti-aligned dipolar excitons, effectively binding them into composite quasiparticle states. By tuning the dipolar excitons into non-degenerate states via the QCSE, we realize *in situ* reconfiguration of the polariton wavefunction and observe an emergent polariton branch exhibiting non-monotonic Stark shifts. Notably, these tunable polaritons allow for customized control over nonlinear interactions through distinct excitonic hybridization and dipolar configurations. This *in situ* tunability offers a scalable pathway toward electrically programmable quantum fluids of light and correlated polariton phases in on-chip photonic integrated circuits.

# Introduction

Interacting photons are a central resource for quantum optics and nonlinear information processing[1,2]. Through strong light–matter coupling in semiconductor cavities, exciton polaritons offer a practical route by combining photonic coherence and transport with exciton-mediated nonlinear interactions[2–6]. These key features have enabled diverse macroscopic quantum phenomena rooted in Bose-Einstein condensation[7,8], and underpin promising on-chip photonic applications [9–13] and quantum simulations[14,15]. Although strategies such as artificial lattice potentials and charged exciton complexes have been employed to strengthen polariton interactions[16–21], the short-range nature of excitonic exchange interactions remains a recurring bottleneck. Dipolar excitons can surmount this limitation via dipolar interactions arising from their out-of-plane (OP) static electric dipole moments. While such capability has been successfully demonstrated with versatile tunability in conventional quantum well systems[22–24], van der Waals (vdW) semiconductors offer a compelling platform[25–34]. Owing to their discrete layered structure, vdW heterostructures host interlayer excitons (IX) with layer-addressable out-of-plane (OP) dipole moments[26,28–30,34], enabling the precise engineering of dipolar interactions. However, they typically face a fundamental trade-off between oscillator strength and interaction strength[25–34], impeding the formation of strongly interacting dipolar polaritons in the strong coupling regime.

Naturally stacked 2H-bilayer $MoS_2$ (2L-$MoS_2$) overcomes this trade-off via tunneling-mediated hybrid IX of dipolar excitons. Facilitated by interlayer hole tunneling, the hybrid IX inherit substantial oscillator strength from intralayer excitons while maintaining strong dipolar interactions with OP dipole moments[35–42]. Consequently, the IX in 2L-$MoS_2$ can achieve strong coupling with cavity photons, forming dipolar polaritons with significantly enhanced nonlinear interactions[43–45]. Crucially, the hybrid IX consist of two degenerate dipolar states with anti-aligned OP dipole moments, capable of simultaneously supporting repulsive and attractive interactions[40] for exploring rich many-body physics[35–42,46]. An applied vertical electric field ($E_z$) lifts this degeneracy via the quantum-confined Stark effect (QCSE)[35–42,46], resulting in two anti-aligned dipolar states characterized by split resonances and reduced oscillator strengths. This dipolar configuration enables on-demand programmability of the anti-aligned dipolar states and their polaritonic interactions. This capability is largely inaccessible to conventional quantum well systems[22–24], and has not yet been experimentally demonstrated.

In this context, we leverage the hybrid dipolar IX to demonstrate electrically tunable dipolar polaritons in a gated 2L-MoS$_2$ device integrated with a one-dimensional photonic crystal (1DPC) hosting bound states in continuum (BIC). In the strong coupling regime, the resulting polaritons hybridize cavity photons with a coherent superposition of two anti-aligned dipolar excitons, effectively binding them into composite quasiparticle states. By applying an $E_z$-field via the QCSE, we lift the IX degeneracy into two distinct anti-aligned dipolar states with split resonances. This enables electrical reconfiguration of the polariton wavefunction via tunable IX hybridization, leading to the emergence of a new polariton branch. The new branch effectively binds these non-degenerate dipolar excitons and exhibits non-monotonic Stark shifts driven by the evolving IX composition. This reconfiguration of dipolar polaritons grants the *in situ* tunability of nonlinear bosonic interactions across different polariton branches. Ultimately, the ability to reconfigure the polariton wavefunction via tunable dipolar IX provides a pathway toward electrically programmable quantum fluids of light and exotic correlated polariton phases.

## Results and discussions

### Characterization of the hybrid 2L-MoS$_2$ system

We developed a hybrid architecture integrating a gated 2L-MoS$_2$ device with a 1DPC for this work, achieving strong coupling between the IX and photonic modes (Fig. 1a; see Methods and Supplementary Notes 1-2 for details). The gating device functions as a planar capacitor, consisting of a bilayer MoS$_2$ encapsulated by an insulator of few-layer hexagonal boron nitride (hBN), and contacted by two few-layer graphene electrodes. This configuration allows for the application of a tunable perpendicular electrostatic $E_z$-field to the 2L-MoS$_2$ to induce the QCSE of dipolar excitons. The gated device is fabricated atop the 1DPC, which features a Si$_3$N$_4$ grating structure with a period of 365 nm, a height of 106 nm, and a filling factor of 0.65 (see Methods and Supplementary Note 1 for details). The 1DPC hosts BIC with an extremely high quality factor (Q-factor) at the Γ point[47]. Crucially, the 1DPC photonic modes exhibit a highly confined electromagnetic field near the interface between Si$_3$N$_4$ grating and the gating device, facilitating the strong coupling between the 1DPC and 2L-MoS$_2$ (see Supplementary Notes 1 and 3, Supplementary Figs. 1-2 for simulation details). With the fabricated samples, Fig. 1b shows a typical microscopic image. The lateral dimension of the exfoliated 2L-MoS$_2$ flake is ~ 40 μm, while the hBN and graphene extend to ~80 μm, ensuring optimal conditions for optical characterization and electrical control. A portion of the 2L-MoS$_2$ flake extends beyond the 1DPC region; this uncoupled area serves as a reference for accurately monitoring the intrinsic tunable exciton resonances.

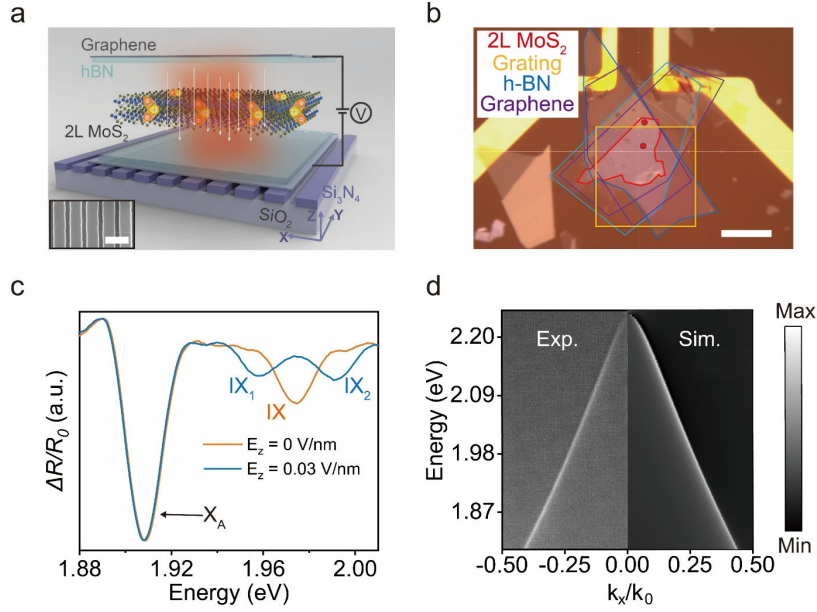

Fig. 1. Hybrid 2L-MoS$_2$ system for the tunable dipolar polaritons. (a) Schematics of a gated 2L-MoS$_2$ device on a Si$_3$N$_4$ grating-based 1DPC. The hybrid dipolar IX in 2L-MoS$_2$ strongly couple with the 1DPC photonic modes and can be controlled by the applied $E_z$-field. The inset shows the SEM image of 1DPC with a scale bar of 500 nm. (b) Optical microscopic image of one hybrid 2L-MoS$_2$ sample. The orange square outlines the area of the 1DPC. The red, blue, and purple lines mark the boundaries of 2L-MoS$_2$, few-layer hBN, and graphene, respectively. The two red dots represent the area on/off the 1DPC. The scale bar is 30 μm. (c) The reflectance contrast spectra of 2L-MoS$_2$ with/without an applied $E_z$-field at 6 K. The hybrid IX is split into IX$_1$ and IX$_2$ with an $E_z$-field. (d) The experimental (left) and simulated (right) photonic dispersions of the 1DPC under TE polarization exhibit excellent agreement.

We then characterized the exciton resonances of the bare 2L-MoS$_2$ and the photonic resonances of the 1DPC. Figure 1c plots the reflectance contrast spectra of bare 2L-MoS$_2$ with/without an applied $E_z$-field. In the bare 2L-MoS$_2$ without $E_z$-field, we identify the main exciton resonance ~ 1.907 eV as A exciton (X$_A$) and the resonance at 1.974 eV as the hybrid IX (Fig. 1a, see Supplementary Note 2 and Supplementary Fig. 3)[36]. In 2L-MoS$_2$, the IX feature localized electrons in one layer and delocalized holes across the bilayer, exhibiting degenerate dipolar states with anti-aligned OP dipole moments ($p_z = \pm e \cdot d$) and possess considerable oscillator strengths[36,37,48]. Under an applied $E_z$-field, the QCSE lifts the dipolar IX degeneracy into two split resonances (IX$_1$ and IX$_2$) as $E_{IX_1} = E_{IX} - E_z \cdot p_z$ and $E_{IX_2} = E_{IX} + E_z \cdot p_z$, where $E_{IX}$, $E_{IX_1}$ and $E_{IX_2}$ represent the exciton resonances of IX, IX$_1$, and IX$_2$, respectively (Supplementary Fig. 4, and Supplementary Note 8)[36,37,48]. For example, at an $E_z$-field of 0.03 V/nm, the resonances of IX$_1$ and IX$_2$ are well resolved

(Fig. 1c). In contrast, the QCSE is negligible for $X_A$, as they do not possess any OP dipole moments. Figure 1d displays the experimental and simulated photonic dispersions of the 1DPC under transverse-electric (TE) polarization (see Methods, Supplementary Note 1 and 3 for more details), which are in close agreement. The broad spectral range and tight confinements of the photonic modes (with a BIC mode ~ 2.24 eV at $\Gamma$ point) enable the efficient coupling with all the excitons of interest in 2L-MoS$_2$.

**Strong coupling regime and electrical tunability**

We performed angle-resolved ($k$-space) reflectance spectroscopy to characterize the strong coupling regime and electrical tunability on the hybrid integrated sample of 2L-MoS$_2$ and 1DPC under different conditions. Figure 2a shows $k_x$-resolved reflectance spectra of the sample without the $E_z$-field. Pronounced anti-crossing features are observed in the vicinity of the IX and $X_A$ resonances, indicating the formation of multiple polariton branches by the strong coupling between different excitons and the shared photonic mode. Here, we focus on the polariton branches associated with strong coupling to the IX, while a detailed discussion of coupling involving other excitons (e.g., $X_A$) is provided in Supplementary Note 4 and Supplementary Fig. 6. Intriguingly, upon applying an $E_z$-field ($E_z$ = 0.035 V/nm), the feature of two polariton branches in the strong coupling regime of the sole IX evolves into two anti-crossing features with two energy-split excitons, leading to three polariton branches with an emergent middle branch (Fig. 2b). This indicates that the photonic mode now strongly couples with two emergent IX under the applied $E_z$-field, i.e., the QCSE induced IX$_1$ and IX$_2$. As the $E_z$-field increases ($E_z$ = 0.05 V/nm), the three polariton branches are separated further, as expected for the strong coupling with more split resonance energies of IX$_1$ and IX$_2$ (Fig. 2c). This suggests the strong coupling with IX$_1$ and IX$_2$ persists even though these two excitons are further apart with increased $E_z$-field. Here, we label the polariton branches as lower, middle, and upper branches (LPB, MPB, UPB) based on their energies, as shown in Fig. 2b and 2c.

We employed a two-coupled-oscillator model to analyze the polariton Hamiltonian in the absence of $E_z$-field as (Supplementary Note 4)[16]:

$$H = \begin{pmatrix} E_{\mathrm{ph}} - i\gamma_{\mathrm{ph}}/2 & V \\ V & E_{\mathrm{IX}} - i\gamma_{\mathrm{IX}}/2 \end{pmatrix} \quad (1)$$

where $E_{ph}$ and $\gamma_{ph}$ represent the energy and linewidths of the photonic mode, $\gamma_{IX}$ represents the IX linewidth, $V$ denotes the exciton-photon coupling strength. By fitting the experimentally extracted polariton resonances (the dashed rectangular area in Fig. 2a), we determine Rabi splitting $\hbar\Omega_{IX} = \sqrt{4V^2 - (\gamma_{ph} - \gamma_{IX})^2} = 38.06 \pm 0.16\ meV$ between the LPB and UPB as shown in Fig. 2d. When the IX$_1$ and IX$_2$ emerge via the QCSE, we analyzed the polariton Hamiltonian via a three-coupled-oscillator model as (Supplementary Notes 3-5, Supplementary Fig. 5)[16]:

$$H = \begin{pmatrix} E_{ph} - \frac{i\gamma_{ph}}{2} & V_{IX_1} & V_{IX_2} \\ V_{IX_1} & E_{IX_1} - \frac{i\gamma_{IX_1}}{2} & 0 \\ V_{IX_2} & 0 & E_{IX_2} - \frac{i\gamma_{IX_2}}{2} \end{pmatrix} \quad (2)$$

where $\gamma_{IX_1}$ and $\gamma_{IX_2}$ represent IX$_1$ and IX$_2$ linewidths, $V_{IX_1}$ and $V_{IX_2}$ denote the coupling strengths for IX$_1$ and IX$_2$, respectively. Similarly, in Fig. 2e, f, we determine the Rabi splittings by fitting the polariton resonances (the rectangular areas in Figs. 2b and 2c) as: $\hbar\Omega_{IX_1} = 30.58 \pm 0.56\ meV$, $\hbar\Omega_{IX_2} = 19.25 \pm 0.56\ meV$ at $E_z = 0.035$ V/nm; $\hbar\Omega_{IX_1} = 27.48 \pm 0.56\ meV$, $\hbar\Omega_{IX_2} = 30.96 \pm 0.56\ meV$ at $E_z = 0.050$ V/nm. Clearly, all the Rabi splittings always satisfy the strong coupling criterion $(\hbar\Omega)^2 > (\gamma_{ph}^2 + \gamma_{IX}^2)/4$, verifying the robust strong coupling regime in all the cases. We reproduced these results in an independent device (Supplementary Fig. 14), demonstrating the robustness and reproducibility of the strong coupling regime and electrical control.

Notably, both $\hbar\Omega_{IX_1}$ and $\hbar\Omega_{IX_2}$ are smaller than the zero-field $\hbar\Omega_{IX}$ as the $E_z$-field increases. Since the Rabi splitting $\hbar\Omega$ is proportional to the square root of the oscillator strength ($\Omega \propto \sqrt{f}$), this reduction indicates that the oscillator strength of the original IX resonance is distributed between the anti-aligned IX$_1$ and IX$_2$ states. This behavior is consistent with the QCSE: the initially degenerate IX splits into two non-degenerate IX$_1$ and IX$_2$, effectively decomposing the IX oscillator strength ($f_{IX}$) into $f_{IX_1}$ and $f_{IX_2}$. Moreover, the extracted $\hbar\Omega_{IX_1}$ and $\hbar\Omega_{IX_2}$ exhibit a trade-off as the $E_z$-field is tuned. This is likely attributed to the opposing $E_z$-field dependences of the anti-aligned dipolar excitons, where the IX$_1$ becomes more polarized by the QCSE accompanied by a decrease in energy and $f_{IX_1}$ while the IX$_2$ exhibits the converse behavior. We also analyzed the strong coupling involving X$_A$ and the corresponding polaritons, which remain essentially insensitive to the applied $E_z$-field (Supplementary Note 4 and Supplementary Fig. 6).

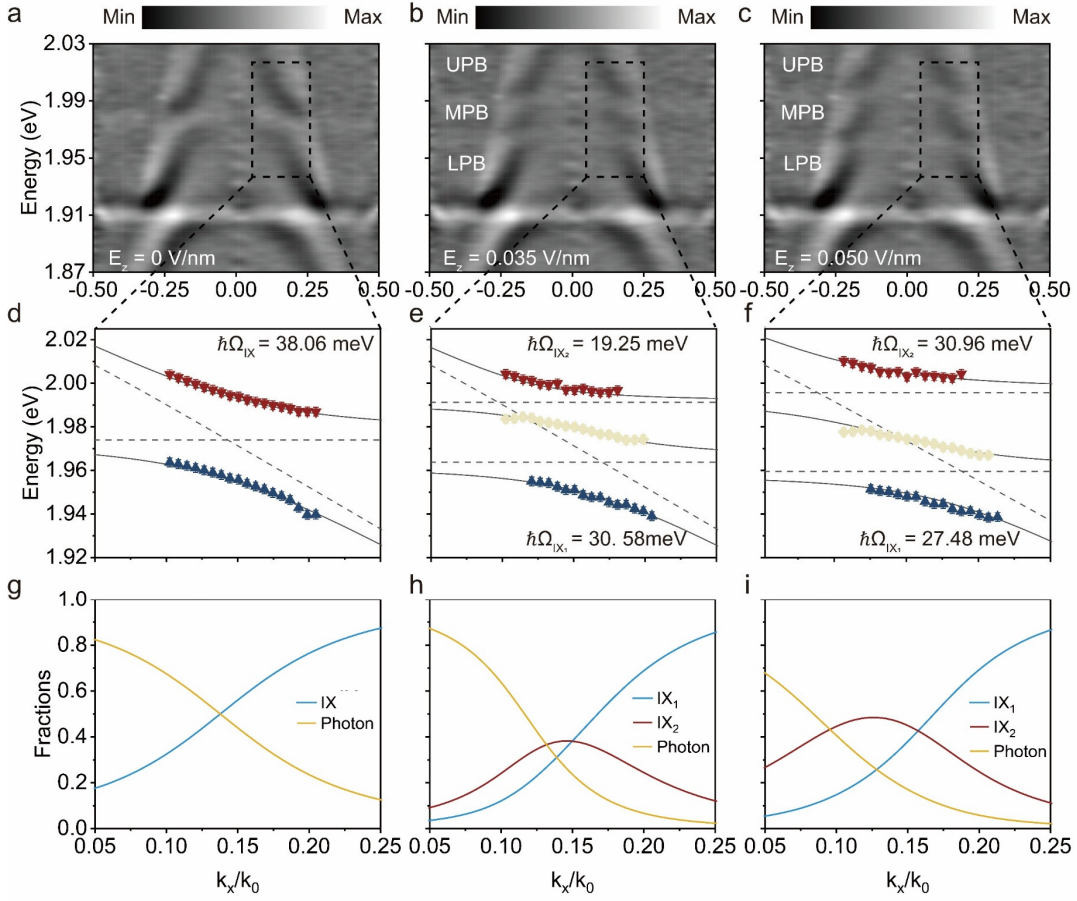

Fig. 2. Tunable strong coupling regime via the QCSE. Angle-resolved ($k_x$-resolved) reflectance spectra (a) at zero $E_z$-field, (b) at $E_z = 0.035$ V/nm, and (c) at $E_z = 0.050$ V/nm. (d-f) Extracted polariton energies (data points) in the zoomed-in rectangular regions from (a-c) and fitted polariton dispersions (solid curves). Note that the dashed lines denote the uncoupled photonic modes and excitonic resonances. (g-i) The constituent fractions derived from Hopfield coefficients (g) at zero $E_z$-field, (h) at $E_z = 0.035$ V/nm, and (i) at $E_z = 0.050$ V/nm.

Leveraging the coupled-oscillator model, we elucidate the field-tunable correlation in the hybridization of anti-aligned dipolar excitons within the strong coupling regime. Figures 2g-i display the corresponding exciton and photon fractions of LPB at zero $E_z$-field and MPB at $E_z = 0.035$ V/nm, $0.050$ V/nm. At zero $E_z$-field, the polariton branches are the hybridization of the degenerate IX resonance and cavity photons. Upon the application of the $E_z$-field, however, all polariton branches evolve into hybrid states composed of photons and both anti-aligned dipolar excitons with field-dependent weighting. While the LPB (UPB) is dominated by the hybridization of IX$_1$ (IX$_2$) and photons (Supplementary Fig. 9), the emergent MPB exhibits a distinct character, primarily composed of a coherent superposition of both IX$_1$ and IX$_2$. Crucially, this distinct MPB effectively mediates the binding of two anti-aligned dipolar excitons into a single composite quasiparticle, as detailed below.

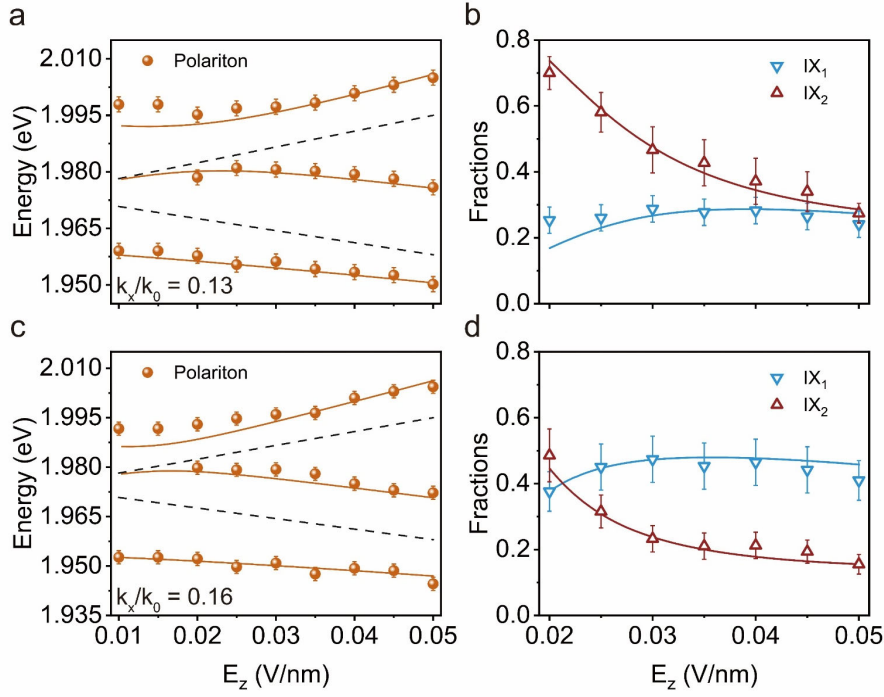

Fig. 3. Tunable hybridization of dipolar excitons. Extracted polariton energies (colored data points) and uncoupled dipolar exciton energies (dashed lines) as a function of the $E_z$-field for the spectra at (a) $k_x/k_0 = 0.13$ and (c) $k_x/k_0 = 0.16$. The solid lines denote the calculated polariton energies with the $E_z$-field-dependent coupled-oscillator model. The excitonic fractions of MPB based on the experimental data (discrete points) and the predicted fractions from the $E_z$-field-dependent coupled-oscillator model for (b) $k_x/k_0 = 0.13$ and (d) $k_x/k_0 = 0.16$.

To elucidate the $E_z$-field-dependent hybridization of the dipolar excitons, we analyze the polaritons at two typical in-plane momenta ($k_x/k_0 = 0.13$ and $0.16$) with $E_z$-field-dependent coupled-oscillator model. Figures 3a and 3c respectively plot the extracted polariton energies at the two momenta, as the $E_z$-field is continuously varied. For reference, we also overlay the field-dependent dipolar energies (dashed lines) of non-degenerate, anti-aligned IX$_1$ and IX$_2$ based on their linear Stark trends of the uncoupled 2L-MoS$_2$ region (Supplementary Note 5 and Supplementary Fig. 7). Correspondingly, although the detailed $E_z$-field dependence differs at the two momenta, the polaritonic spectrum in both cases evolves from two branches (UPB and LPB) to three branches (UPB, MPB, and LPB). The UPB (LPB) exhibits monotonic blueshifts (redshifts), reflecting the dominant contribution from the blue-shifted $IX_2$ (red-shifted $IX_1$) induced by the QCSE. Crucially, MPB resides energetically between $IX_1$ and $IX_2$ states and exhibits non-monotonic Stark shifts, arising from a distinct superposition state that effectively binds both anti-aligned dipolar excitons.

We further interpret the distinctive hybridization of these polariton branches using an $E_z$-field-dependent coupled-oscillator model. For each value of $E_z$, the excitonic fractions (from the Hopfield coefficients) are obtained by fitting the calculated polariton energies to the experimentally extracted data, as shown by the solid curves in Figs. 3a and 3c (see Supplementary Note 5 and Supplementary Fig. 8 for details). The resulting $IX_1$ and $IX_2$ excitonic fractions of the MPB are plotted in Figs. 3b and 3d, and consistent with the experimentally extracted fractions (overlaid scatter points). Here, we focus on the hybridization of the anti-aligned dipolar excitons; the associated field-dependent photon fractions are presented in Supplementary Fig. 10. Specifically at $k_x/k_0 = 0.13$, the exciton fractions of the $IX_1$ and $IX_2$ in the MPB are strongly imbalanced at small $E_z$-field but are more balanced at large $E_z$-field. On the contrary, at $k_x/k_0 = 0.16$, the exciton fractions of $IX_1$ and $IX_2$ are almost balanced at small $E_z$-field but become increasingly imbalanced at large $E_z$-field. Despite these momentum-dependent variations, the MPB consistently hybridizes photons with a coherent superposition of the anti-aligned, non-degenerate $IX_1$ and $IX_2$. Consequently, the MPB effectively bind both anti-aligned IX with field-tunable hybridization weights that can be widely and reversibly reconfigured by the QCSE, resulting in non-monotonic Stark shifts. As the $E_z$-field increases, the MPB at $k_x/k_0 = 0.13$ initially exhibits a slight blueshift due to the dominant contribution of the blue-shifted $IX_2$, eventually returning to zero-field energy as the contributions from the two IX become balanced. Conversely, at $k_x/k_0 = 0.16$, the MPB initially displays a negligible Stark shift due to balanced IX contributions, but subsequently undergoes a redshift as the $IX_1$ contribution becomes dominant.

**Polariton nonlinearity controlled by tunable hybridization of dipolar excitons**

By exploiting the electrically tunable hybridization of dipolar excitons, we investigated polariton nonlinearity using power-dependent spectroscopy and examined how the nonlinear response correlates with the constituents. Specifically, we analyze the power-dependent reflectance spectral shifts of representative polariton branches (i.e., LPB-$X_A$, LPB-IX, LPB-$IX_1$, UPB-$IX_2$, MPB) at selected $k_x/k_0$ and $E_z$-fields to disentangle nonlinear contributions associated with dominant $X_A$, IX, $IX_1$, and $IX_2$ constituents (See details in Supplementary Note 6 and Supplementary Figs.

11-12). As the polariton density increases, LPB-$X_A$ (with a dominant $X_A$ fraction of 0.5) exhibits slight blueshifts regardless of the applied $E_z$-field (Fig. 4a and Supplementary Fig. 11), which is similar to previous report[16]. However, the LPB-IX (with a dominant IX fraction of 0.5) undergoes apparent blueshifts and exhibits significantly stronger nonlinearity than LPB-$X_A$ (Fig. 4b). When $E_z = 0.025$ V/nm is applied, LPB-$IX_1$ (with a dominant $IX_1$ fraction of 0.6) and UPB-$IX_2$ (with a dominant $IX_2$ fraction of 0.5) show apparent blueshifts and redshifts, respectively, demonstrating a convergence trend towards the $IX_1$ and $IX_2$ states (Fig. 4c). Intriguingly, we observe complex saturation-like redshifts for the MPB (characterized by $IX_1$ and $IX_2$ fractions of 0.2, 0.5, respectively), signaling unique nonlinear interactions via the dipolar exciton hybridization.

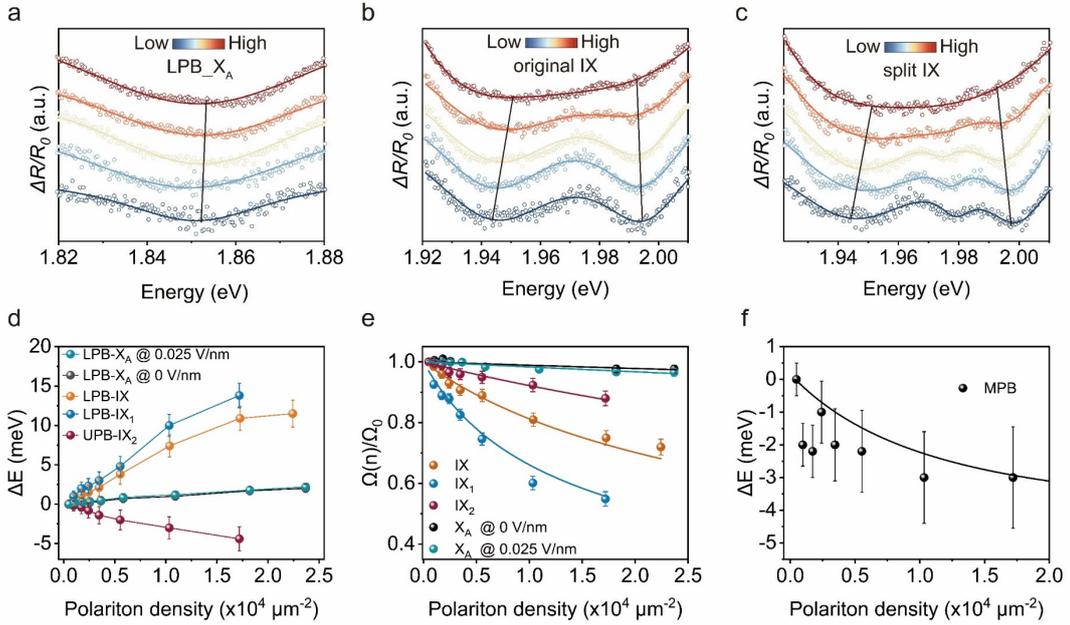

Fig. 4. Distinct polariton nonlinearity and quantitative analysis. Reflectance contrast spectra of (a) around LPB-$X_A$ at zero $E_z$-field, (b) around IX at zero $E_z$-field, and (c) around $IX_1$ and $IX_2$ at $E_z = 0.025$ V/nm as a function of pump fluence. Solid lines in (a-c) are fitting results by the Lorentzian model. Summarized (d) energy shifts for representative polariton branches, (e) Rabi splittings, and (f) MPB energy shifts as a function of the polariton density. The solid lines in (e) are theoretical fits, and the solid curve in (f) is the calculated energy shift.

To quantitatively assess the exciton-dependent polariton nonlinearity, we extract the density-induced energy shifts of the selected branches (Fig. 4d) and compare them with the power-dependent Rabi splittings (Fig. 4e). The summarized shifts in Fig. 4d indicate that the polariton interaction strength is weakest for the LPB-$X_A$, significantly enhanced for the LPB-IX, and strongest for the LPB-$IX_1$. In contrast, the LPB-$IX_2$ exhibits a moderate interaction strength with the opposite sign (see

Supplementary Note 6 and Supplementary Fig. 11-13 for more details). By analyzing the polariton hybridization and power-dependent Rabi splittings (Fig. 4e), we find that the polariton nonlinearity in all these representative branches arises primarily from the phase-space filling (PSF) effect. Generally, polariton nonlinearity originates from two mechanisms: PSF-induced saturation and excitonic interactions. To quantify the PSF contribution, we extracted the power-dependent Rabi splitting for $X_A$, IX, $IX_1$, and $IX_2$ (Fig. 4e). The $X_A$ Rabi splitting exhibits only a minor reduction, consistent with the tiny spectral blueshifts for LPB-$X_A$. In contrast, the IX Rabi splitting shows a pronounced attenuation with increasing density. Moreover, the attenuation occurs more rapidly for $IX_1$ compared to $IX_2$, revealing distinct saturation behaviors. These trends and the energy shifts are in good agreement with the branch-dependent spectral shifts discussed above, suggesting the polariton nonlinearity is dominated by the reduction of Rabi splitting via PSF effect. Furthermore, we estimated the expected energy shifts based on the exciton fractions of each polariton branch and bare exciton interactions[40], which are significantly smaller than the observed polariton energy shifts. This confirms that the polariton nonlinearity predominantly stems from the PSF effect.

To elaborate on the PSF effect, we adopt a simplified model commonly used to describe PSF-induced saturation[16] as $\Omega(n) = \frac{\Omega_0}{\sqrt{1+\frac{n}{n_s}}}$, where $n_s$ denotes the saturation density. We extracted the saturation densities for $X_A$, IX, $IX_1$, and $IX_2$, which are respectively $4.43 \times 10^5 \ \mu m^{-2}$ ($3.01 \times 10^5 \mu m^{-2}$ at $E_z$ = 0.025 V/nm), $1.94 \times 10^4 \ um^{-2}$, $7.76 \times 10^3 \ um^{-2}$, and $5.56 \times 10^4 \ \mu m^{-2}$. While the saturation density of $X_A$ is much higher than that of IX, $IX_1$, and $IX_2$. The difference between $IX_1$ and $IX_2$ likely originates from the QCSE-modified dipolar exciton wavefunctions[41]. Particularly, the applied $E_z$-field enhances the IX spatial distribution and interaction strength of the $IX_1$ (and thus earlier saturation), whereas the opposite trend applies to $IX_2$.

With the established PSF mechanism, we can further understand the spectral redshifts observed in the MPB (Fig. 4f). By applying the saturation model to the hybridized anti-aligned $IX_1$ and $IX_2$ (Supplementary Note 6)[13,43], we approximate the expected energy shifts, which show qualitative agreement with our experimental results. Although precise quantification of the polariton nonlinearity warrants further investigation using higher-quality samples with more accurate measurements, this analysis suggests that the MPB nonlinearity arises from the synergistic interplay of the composite anti-aligned dipolar excitons.

## Conclusion

In this work, we demonstrate electrically programmable strong coupling and distinct nonlinear interactions of dipolar polaritons with tunable IX hybridization in a 2L-MoS$_2$–1DPC system. The tightly confined photonic mode enables strong coupling to bright dipolar excitons under the influence of the QCSE within a single planar architecture. Utilizing the electrically compatible grating geometry, we apply a vertical electrostatic field across the 2L-MoS$_2$ to lift the IX degeneracy and Stark-tune the anti-aligned dipolar excitons, thereby continuously reconfiguring the polariton composition. As the split IX$_1$ and IX$_2$ emerge, they hybridize with the cavity photon to form an emergent MPB, characterized by non-monotonic Stark shifts with electrically addressable anti-aligned dipolar configurations. Crucially, the applied field serves as an *in situ* control knob to program the dipolar configurations, exciton hybridization, and consequently the nonlinear response of the polaritons. Looking forward, integrating this tunable platform with moiré superlattices may enable flat-band and topological polariton engineering, while further enhancing photon confinement could drive the system toward the polariton-blockade[49,50] regime for scalable quantum photonic devices.

# Methods

## Sample preparation

**Si$_3$N$_4$ grating with integrated electrodes.** The Si$_3$N$_4$ film was first evaporated via PECVD on a pre-clean silica substrate, followed by electron beam lithography and reactive ion etching to obtain the grating structure. The grating was characterized by scanning electron microscope and angle-resolved reflection spectra, and the applicable gratings were adopted for optical lithography to fabricate the electrodes.
**TMD heterostructures.** The heterostructure was constructed from mechanically exfoliated 2H-stacked bilayer MoS$_2$, hBN, and graphene flakes. The bilayer MoS$_2$ was first identified by optical microscopy and further confirmed by photoluminescence spectroscopy, as well as its characteristic reflectance. Few-layer hBN and graphene flakes were prepared using the same exfoliation method. To minimize optical absorption and cavity loss, the graphene was chosen as thin as possible. The stack was assembled in a graphene/hBN/bilayer MoS$_2$/hBN/graphene sequence using a PDMS/polycarbonate film-based dry-transfer process, and then transferred onto the pre-fabricated grating with pre-patterned electrodes, with the graphene gates precisely aligned to the contacts. Finally, the device was annealed at 200 °C for 2h under a nitrogen atmosphere to improve interfacial flatness and remove residual organics.

## Optical Characterization

All optical experiments were performed at 6 K in a Montana cryostat. The optical setup was a home-built Fourier-imaging microscope capable of switching between real-space and momentum-space measurements. Excitation and collection were both realized using a 50× objective (NA = 0.5) integrated into an Olympus microscope. All signals were dispersed by a spectrometer (Princeton Instruments, 600 lines/mm grating) and detected by a charge-coupled device.
For steady-state reflection spectroscopy, a broadband halogen lamp was used as the light source. For nonlinear optical experiments, a broadband supercontinuum white-light source was generated by focusing a narrowband femtosecond laser (50 kHz repetition rate, 222 fs pulse duration) onto a sapphire crystal. The bias voltage was applied using a source meter.

## Simulation

All the simulation results were obtained based on COMSOL Multiphysics 6.3. We set the material parameters that are consistent with the experiments. These parameters include the refractive index of silicon nitride, the structure geometry parameters of the grating, the refractive indices, and the actual thicknesses of each layer in the hybrid vdW heterostructure of hBN-encapsulated 2L-MoS$_2$.

## Acknowledgments

This work was supported by the National Key Research and Development Program of China (No. 2021YFA1200800), the National Natural Science Foundation of China (Grant Nos. 12304429, 62261160386), and the Open Research Fund of the Pulsed High Magnetic Field Facility (Grant No. WHMFC2024015), Huazhong University of Science and Technology.
We thank Dr. Xinyuan Zhang from the Core Facility of Wuhan University for her assistance with AFM (atomic force microscope) test.

## Author contributions

X.L. Q.G, and H.K. conceived and designed the project. H.K. carried out all optical spectroscopy measurements and characterized the devices, and prepared the samples with assistance from T.Z., W.D. and Q.G. H.K. analyzed the data with input from X.L. and Q.G. T.T. and K.W. provided high-quality hBN crystals. F.S., T.Y. and H.X. provided important guidance on manuscript preparation. All authors contributed to the paper. All work was supervised by X.L., Q.G., and T.Y.

## Conflict of interest

The authors have no conflicts to disclose.